# WORLD UNIVERSITY RANKINGS – A PRINCIPAL COMPONENT ANALYSIS*

*Joao E. Steiner*

*Instituto de Estudos Avançados*
*Instituto de Astronomia, Geofísica e Ciências Atmosféricas*
*Universidade de São Paulo*
(steiner@usp.br)

**Abstract**

In order to establish which parameters and corresponding weights are more appropriate for the assessment of academic excellence in the context of university rankings, I have made a multivariate data analysis on a set of 13 parameters for 178 institutions. I found that the three more relevant components are academic excellence (48%), internationalism (14%) and faculty/student ratio (8%). It is shown that these components are not correlated.

# Introduction

Institutions are, in general, in permanent quest for excellence. But… in the case of universities, what is really excellence?  The word somehow expresses a diffuse idea, given its complexity but also the universality and scope of the involved subjects. How can one combine themes of humanities with those of natural sciences?   How can one equilibrate pure and applied studies? We will hardly find a singular parameter that can evaluate and express, in a credible form, subjects that are so complex.

The evaluation of universities has been the theme of recurrent discussions. Some of these evaluations have been expressed in the form of rankings. University rankings have been frequent among American and Canadian universities, for example. In recent years, some initiatives have established world rankings. The visibility of such initiatives may have implications on policies of institutional development. "Rankings serve a variety of purposes, good and bad … and are also inevitable in the era of massification" (Altbach, 2006).

 As there is not a singular parameter that expresses the university excellence, the rankings have been based on a number of parameters that focus the institutional performance in a variety of perspectives. The larger the number of parameters and the better they are selected, greater are the chances that the conclusions will have credibility. A small number of parameters (would 6 be reasonable?) will always be regarded as problematic. In addition, when dealing with institutions of a large number of countries, uniformity becomes an issue (van Raan, 2005).  One faces the problem of heterogeneity of criteria, languages and cultures. Even more problematic may be the form of establishing weights for the distinct parameters.

The choices of both, parameters and weights, may represent cultural, political or economic perspectives that could introduce non-universal values and, therefore, should be regarded with caution.

*Existing world university rankings*

In recent years, three rankings of world universities have been published.  We will list, here, the main parameters and respective weights (in parenthesis). A label from "A" to "M" has been assigned to each parameter according to Table 3.

- **Shanghai Jiao Tong University ranking - SJTU.** This ranking is established using six parameters and arbitrarily attributed weights.

A -    (20%) -  N&S. Number of articles published in Nature or Science;

B -    (20%) -  HiCi. Number of highly cited researchers in 21 broad subject categories;

C -    (10%) -  Size. Academic performance with respect to the size of the institution;

D -    (20%) -  Award. Number of staff of the institution winning Nobel Prize and Field Medals;

E -    (10%) -  Alumni. Number of alumni of the institution winning Nobel Prizes and Fields Medals;

F -    (20%) -  SCI. Articles in Science Citation Index-expanded, Social Science Citation Index, and Arts and Humanities Citation Index;

- The ranking of the **Times Higher Education Supplement** - **THES** adopted the following parameters and respective weights:

G -   (40%) -   Peer review – opinion of 2,375 research-active academics;

H -   (20%) -   Citation/faculty;

I -   (10%) -   Recruiter's review – the opinion of employers;

K -   (20%) -   Faculty to student ratio;

L -   (5%) -    International student score – percentage of foreign students;

M -   (5%) -    International faculty score – percentage of foreign staff.

- The **Webometrics** ranking measures the presence of the universities in the web, considering parameters such as size of the sites, rich files and visibility.
J -    Presence in the web.

While the THES ranking is based on relative numbers such as indices, ratios etc, the SJTU mixes absolute numbers such as number of papers, scientists and prizes with relative numbers such as academic performance relative to size. The Webometrics ranking is established in terms of absolute numbers. It is important to notice that in one case we look at the relative performance, regardless of the size, while in the other case the size of the institution also impacts the ranking. In this case, larger institutions tend to be better ranked than smaller ones.

The present study performs an evaluation with a larger number of parameters by combining the 13 parameters described above. The main goal is to find out which parameters are more and less important for the assessment of the academic performance in the world universities. The focus is to discuss methodology and not to establish a new ranking.

We utilized the data from universities considered in the 2005 versions of the three rankings mentioned above: SJTU, THES and Webometrics. We found a total of 178 universities that are common among these three studies. The limiting factor of this set is the reduced number of universities in the THES survey that published data for only 200 universities.

The cross-correlation of parameters can be seen in Table 1. Parameters labeled from A to G are highly correlated among themselves (>0.50). Parameters K, L and M are weakly or not correlated at all or even anti-correlated with other parameters. The parameters from A to G are usually taken as canonical indicators of academic performance. It becomes apparent that three parameters (K, L and M), half of the ones considered by the THES, are indicators of aspects unrelated to academic performance.

Table 1 – Cross-correlation of parameters

| | A | B | C | D | E | F | G | H | I | J | K | L | M |
|---|---|---|---|---|---|---|---|---|---|---|---|---|---|
| *A-Nat&Sci* | 1 | | | | | | | | | | | | |
| *B-HiCi* | 0,90 | 1 | | | | | | | | | | | |
| *C-Size* | 0,83 | 0,78 | 1 | | | | | | | | | | |
| *D-Award* | 0,78 | 0,72 | 0,81 | 1 | | | | | | | | | |
| *E-Alumni* | 0,74 | 0,68 | 0,72 | 0,80 | 1 | | | | | | | | |
| *F- SCI-Articles* | 0,74 | 0,72 | 0,61 | 0,52 | 0,58 | 1 | | | | | | | |
| *G-Peer Rev* | 0,60 | 0,55 | 0,57 | 0,62 | 0,63 | 0,59 | 1 | | | | | | |
| *H-Cit/Fac* | 0,71 | 0,68 | 0,71 | 0,60 | 0,46 | 0,42 | 0,35 | 1 | | | | | |
| *I-Recruiter* | 0,49 | 0,51 | 0,47 | 0,50 | 0,49 | 0,36 | 0,57 | 0,31 | 1 | | | | |
| *J-(-)Web* | 0,38 | 0,41 | 0,31 | 0,22 | 0,23 | 0,40 | 0,24 | 0,25 | 0,23 | 1 | | | |
| *K-Fac/Stu* | 0,21 | 0,15 | 0,25 | 0,12 | 0,13 | 0,04 | 0,02 | 0,04 | 0,09 | -0,05 | 1 | | |
| *L-Int'l student* | 0,01 | 0,06 | 0,14 | 0,11 | 0,13 | -0,21 | 0,19 | -0,02 | 0,29 | -0,05 | 0,12 | 1 | |
| *M-Int'l faculty* | -0,21 | -0,21 | -0,04 | -0,09 | -0,12 | -0,27 | 0,07 | -0,24 | 0,06 | -0,06 | 0,04 | 0,60 | 1 |

# Principal Component Analysis

In this study we will adopt a distinct strategy for analyzing the question: a multivariate analysis on the form of Principal Component Analysis (Murtag and Heck, 1987). This strategy is more robust than the existing methodology of rankings for two reasons: It does not establish weights a priori. In addition, we consider 13 parameters instead of only 6. This analysis calculates the principal components that maximize the explanation of the variances. If, by hypothesis, the first principal component can be associated to academic performance, then the correlation of the parameters (columns) with respect to this component will establish a relative scale of weights, not a priori but as a result. The correlation of the objects (universities) will establish a scale of performance (ranking) of the universities. For those interested in more details, see the technical note at the end of the paper.

*The Academic performance – Principal Component 1*

Principal Component 1 explains about half (47%) of the data variance (see Table 2). It is strongly determined by the correlation among the columns A, B….J (see Table 3). The top parameters in this correlation are the number of papers published in Nature and Science and the number of highly cited scientists. Such parameters are clearly associated to what one usually understands as academic performance of the institutions. This seems also to be confirmed by the correlation of the objects (universities) with the first principal component (see Appendix A) as most of the universities best placed in this order are also well ranked in the various studies.

Perhaps a surprise is to find that academic performance is not correlated with the internationalism of students (0.10) and faculty (-0.16). The later is even anti-correlated. The ratio faculty/ student is also only weekly correlated with academic performance.

*Internationalism – Principal Component 2*

Principal Components 2 and 3 are, by construction, not correlated with each other and also with respect to the principal component 1; they are responsible for explaining 14% and 8% of the data variance. Principal Component 2 is dominated by the internationalism of the universities as defined by the parameters L and M of Table 3. The correlation with the faculty score is 0.84 and with student score, 0.88. One of the relevant results of the present study is to show that the internationalism is not correlated with the academic performance. The internationalism is weakly correlated with peer review (0.23) and recruiter's review (0.35) and weakly anti-correlated with the number of articles in the Science Citation Index (-0.28).

Principal Component 2, the internationalism of the universities, points towards a geographic discrimination. Appendix 2 shows the 10 first and the 10 last institutions projected on this eigenvector. Among the top ten are universities from UK (4), Switzerland (3), Singapore (2) and France (1). Highly centered in Europe and/or small countries. In the other extreme are universities from large countries such as USA (6) and Brazil (1) or countries that are geographically or politically isolated as Israel (1) and Taiwan (1).

Table 2: Explained Variances

|     | % of Variance | Cumulative % |
|-----|---------------|--------------|
| PC-1 | 47.8 | 47.8 |
| PC-2 | 14.4 | 62.2 |
| PC-3 | 8.4 | 70.6 |
| PC-4 | 6.6 | 77.2 |
| PC-5 | 5.9 | 83.1 |

It seems obvious that to cross the frontier from Germany to Switzerland or to go from Amsterdam to London is easier than to go from New York to California or from Recife to Porto Alegre. Why should the assessment of university academic performance be affected by its locations? Principal Component 2 may perhaps be more useful for geopolitical studies than for university academic rankings.

Table 3 – Correlation of parameters with Principal Components 1, 2 and 3

| Parameter | PC-1 | PC-2 | PC-3 |
|---|---|---|---|
| **A – Articles published in Nature and Science** | 0,94 | -0,10 | 0,09 |
| **B – Highly cited researchers** | 0,91 | -0,09 | 0,02 |
| **C – Academic performance with respect to the size** | 0,89 | 0,08 | 0,17 |
| **D – Staff Nobel/Fields Prizes** | 0,86 | 0,09 | 0,09 |
| **E – Alumni Nobel/Fields Prizes** | 0,83 | 0,09 | 0,03 |
| **F – Articles in the Science Citation Index** | 0,77 | -0,30 | -0,16 |
| **G – Peer review** | 0,73 | 0,24 | -0,25 |
| **H – Citations/faculty** | 0,72 | -0,18 | 0,13 |
| **I – Recruiter review** | 0,62 | 0,34 | -0,18 |
| **J – Presence in the web** | 0,42 | -0,16 | -0,46 |
| **K – Faculty/student ratio** | 0,17 | 0,20 | 0,82 |
| **L – International student score** | 0,08 | 0,89 | -0,03 |
| **M – International faculty score** | -0,16 | 0,83 | -0,15 |

International cooperation is important in science as well as in academic life in general and should not bee confused with the definition of internationalism considered here. The point is how to quantify it. Joint projects, interchange of faculty and students, learning of foreign languages, joint publications etc are of great importance in the promotion of excellence. There are numerous examples on how careful selection of foreign professors has played a strategic role and positively impacted the institutional development. One form of quantifying the international cooperation could, perhaps, be the measurement of multinational co-authorship in publications.

*The faculty/student ratio – Principal Component 3*

Principal Component 3 is dominated by the faculty/student ratio (0.89) and weakly correlated with performance/size (0.21) and anti-correlated with peer review (-0.33), recruiter's review (-0.18) and presence in the web (-0.22).
The extremes of the projections of universities with respect to eigenvector 3 are (see Appendix C): at the high faculty/student ratio, universities from France (5), US (3), Switzerland (1) and Denmark (1); at the low faculty/student ratio, are universities from Australia (5), Canada (2), Singapore (1), US (1) and UK (1). As mentioned above, this component is not correlated with academic performance but it could be relevant for the study of cost/benefits of the higher learning institutions.

*Public versus Private Universities: internationalism and faculty/student ratio*

The international faculty and student score as defined by the THES considers that the more international members in the university, the better it should be ranked. Is this reasonable? In Figure 1 we display the degree of internationalism (PC-2) versus the academic performance (PC-1) only for US universities. We separate public universities from private (not-for-profit) ones. It is clear that they have distinct behaviors. Private universities show a correlation in the sense that institutions with better academic performance have higher degree of internationalism. Public universities show the opposite behavior: universities with higher academic performance have less internationalism. In fact, considering only the institutions ranked better then 40 in academic performance, one can see a clear separation of these two groups. Why is that so? It is clear that private universities look for student from everywhere, as long they pay their (usually high) fees and tuition. Public universities are usually subsidized to some level. Why should they subsidize foreign students? The dichotomy of these two groups, clearly seen in Figure 1 is, therefore, easy to understand. The conclusion is that introducing criteria of international score as an evaluation parameter discriminates unfavorably public universities. If this is a relevant conclusion for a country like the United States, it might be even more relevant for developing countries.
The faculty/student ratio is also correlated with the private versus public nature of the universities. For example, among the 10 US universities ranked highest in PC-3, 9 are private and 1 is public. Among the 10 lowest ranked institutions, all are public. The conclusion is that using this parameter for evaluating academic performance, favors improperly private institutions. As the US News ranking also uses the faculty/student ratio as an evaluation criterion, it is not a surprise that it ranks 20 out of the first 21 universities as private institutions (www.usnews.com/usnews/edu/college/ranking). Appendix A shows that among the 20 highest ranked US universities, 12 are private and 8 are public – which shows much more equilibrated situation.

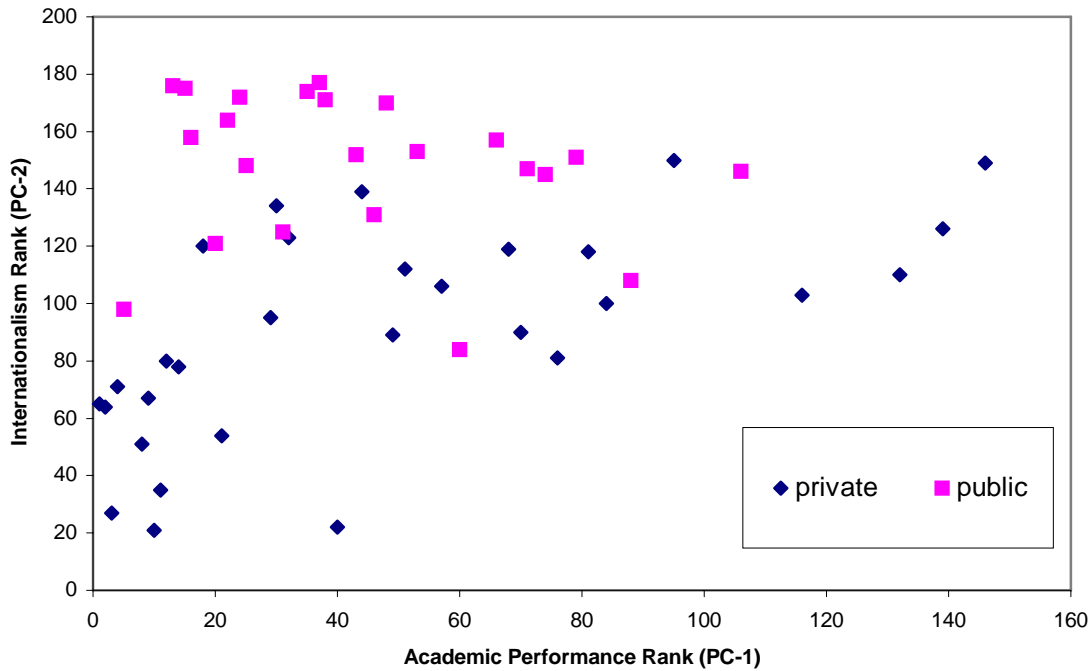

Figure 1 – *The Internationalism (PC-2) versus the Academic Performance (PC-1) for US universities. This displays distinct behaviors between public and private institutions.*

All three rankings analyzed in this study present some flaws. From the six parameters of the THES ranking, only three are correlated with academic performance. The SJTU ranking also is based on six parameters. All six are correlated with PC-1. However five of them are absolute parameters and, in this sense, tend to favor larger institutions. A second flaw is that the Nobel Prize seems to be overweighed. Although both staff and alumni achievements are well correlated with PC-1, the weight attributed to them (total of 30%) seems out of proportion. Finally the correlation of the Webometric ranking with PC-1 is meddling.

## Conclusions

1. A multivariate analysis of a set of indicators for 178 world universities shows that 70% of the data variance can be explained by three main components: the academic performance (Principal Component 1 explains 48% of the variance); the degree of internationalism (Principal Component 2 explains 14% of the variance) and the faculty-to-student ratio (Principal Component 3 explains 8% of the variance).
2. As it is usual in the assessment of institutions, the academic performance is strongly correlated with publications, citations, awards and reviews – canonical indicators of excellence.
3. The degree of internationalism as defined by the THES (this should not be confused with international cooperation in general) is not correlated with academic performance. Distinct countries and regions have different performance with respect to the internationalism, depending on their size, integration with other countries or political and geographical isolation.
4. The ratio faculty/student is only weakly correlated with academic performance. This ratio has also distinct values in different countries, France and Australia being the two extremes.
5. The correlations of the parameters with the three main principal components are provided.
6. A list of universities ordered according to Principal Component 1 is provided. This should not be regarded as a new ranking.
7. Internationalism criteria as well as the faculty/student ratio discriminate unfavorably public universities when compared to private ones.
8. The conclusion derived in the present study may be relevant for institutional strategic planning or for the formulation of public policies. The quest for identifying and implementing policies for supporting world class universities is associated to risks; some of them are identified in this work.

***Technical note***: *The Principal Component Analysis- PCA - is a multivariate procedure in which a set of correlated variables is transformed into a set of uncorrelated variables (called Principal Components) that are ordered by reducing variability (Murtag and Heck, 1987). The uncorrelated variables are a linear combination of the original variables. The principal components are calculated as eigenvectors which, by construction, are orthogonal among themselves and, therefore, uncorrelated. The significance of each eigenvector is expressed as its eigenvalue. The first Principal Component is the combination of variables that explains the greatest amount of variance. The main use of the PCA is to reduce the dimensionality of the data set while retaining as much information as possible. It computes a compact and optimal description of the data set.*

*In the present case, the projection of each column (parameter) with respect to each eigenvector is given in Table 3 for the first three components. The rank of the universities is given in order to the projection of each object with respect to the respective eigenvector (see Appendices).*

## *References*

Altbach, P., *The Dilemmas of Ranking* in: International Higher Education, Number 42, Winter 2006.
Murtag, F. and Heck, A. 1987, *Multivariate Data Analysis*, Reidel Publishing
        Company, Dordrecht, Holland.
SJTU – (http://ed.sjtu.edu.cn/rank/2005)
THES – (www.thes.co.uk)
van Raan, A. F. J., *Fatal attraction: Conceptual and methodological problems in the ranking of universities by bibliometric methods,* 2005, Scientometrics 62, 133
Webometrics – (www.webometrics.info)
USNews - (www.usnews.com/usnews/edu/college/ranking).

*Acknowledgements* – I would like to express my gratitude to Aziz Salem for his technical assistance and to Dr. Francisco Jablonski and to Mauro Bellesa for critical reading of the manuscript.

# Appendix A

## *Academic Performance*

The projection of the objects (universities) according to Principal Component 1. This should not be regarded as a new ranking as the universe of objects was limited to the set of universities in common among the three rankings considered in this paper. The main limiting factor is the small number (200) of universities in the THES rank.

| Rank PC-1 | Rank THES | Rank SJTU | Rank WEB | Name | Country | Projection |
|---|---|---|---|---|---|---|
| 1 | 1 | 1 | 4 | Harvard University | US | 12,26 |
| 2 | 5 | 3 | 3 | Stanford University | US | 8,54 |
| 3 | 2 | 5 | 1 | Massachusetts Institute of Technology | US | 8,12 |
| 4 | 8 | 6 | 42 | California Institute of Technology | US | 8,10 |
| 5 | 6 | 4 | 2 | University of California, Berkeley | US | 7,89 |
| 6 | 3 | 2 | 21 | Cambridge University | UK | 7,86 |
| 7 | 4 | 10 | 28 | Oxford University | UK | 6,02 |
| 8 | 9 | 8 | 39 | Princeton University | US | 5,67 |
| 9 | 20 | 7 | 13 | Columbia University | US | 5,33 |
| 10 | 7 | 11 | 29 | Yale University | US | 5,09 |
| 11 | 17= | 9 | 19 | University of Chicago | US | 4,89 |
| 12 | 14 | 12 | 5 | Cornell University | US | 4,74 |
| 13 | 42 | 13 | 33 | University of California, San Diego | US | 3,73 |
| 14 | 32 | 15 | 14 | Pennsylvania University | US | 3,49 |
| 15 | 37 | 14 | 16 | University of California, Los Angeles | US | 3,33 |
| 16 | 17= | 18 | 99 | University of California, San Francisco | US | 3,31 |
| 17 | 16 | 20 | 83 | Tokyo University | Japan | 3,25 |
| 18 | 27 | 19 | 44 | Johns Hopkins University | US | 3,21 |
| 19 | 29 | 24 | 25 | University of Toronto | Canada | 2,90 |
| 20 | 36 | 21 | 9 | University of Michigan | US | 2,89 |
| 21 | 11= | 32 | 32 | Duke University | US | 2,66 |
| 22 | 73= | 16 | 10 | University Wisconsin-Madison | US | 2,54 |
| 23 | 13 | 23 | 102 | Imperial College London | UK | 2,53 |
| 24 | 88= | 17 | 8 | Washington University | US | 2,40 |
| 25 | 26 | 36 | 7 | University of Texas at Austin | US | 2,36 |
| 26 | 28 | 26 | 68 | University College London | UK | 2,35 |
| 27 | 21 | 27 | 41 | ETH Zurich | Switzerland | 2,28 |
| 28 | 31 | 22 | 190 | Kyoto University | Japan | 2,26 |
| 29 | 46 | 31 | 77 | Northwestern University | US | 2,16 |
| 30 | 58= | 28 | 45 | Washington University, St Louis | US | 2,06 |
| 31 | 58= | 25 | 6 | University of Illinois | US | 1,90 |
| 32 | 56 | 29 | 38 | New York University | US | 1,44 |
| 33 | 30 | 47 | 54 | Edinburgh University | UK | 1,36 |
| 34 | 38= | 37 | 49 | University of British Columbia | Canada | 1,35 |
| 35 | 150= | 32 | 11 | University of Minnesota | US | 1,27 |
| 36 | 24= | 67 | 80 | McGill University | Canada | 1,15 |
| 37 | 64 | 39 | 12 | Pennsylvania State University | US | 1,14 |
| 38 | 159= | 34 | 47 | University of California, Santa Barbara | US | 1,04 |
| 39 | 23 | 56 | 62 | Australian National University | Australia | 0,92 |
| 40 | 44 | 54 | 15 | Carnegie Mellon University | US | 0,89 |
| 41 | 55 | 51 | 199 | Munich University | Germany | 0,85 |
| 42 | 35 | 53 | 582 | Manchester University & Umist | UK | 0,82 |
| 43 | 163 | 35 | 48 | Colorado University | US | 0,77 |
| 44 | 114= | 39 | 136 | Vanderbilt University | US | 0,70 |
| 45 | 49 | 64 | 134 | Bristol University | UK | 0,64 |
| 46 | 133= | 47 | 22 | Maryland University | US | 0,59 |
| 47 | 105= | 52 | 194 | Technical University Munich | Germany | 0,52 |
| 48 | 143= | 55 | 18 | North Carolina University | US | 0,51 |
| 49 | 73= | 65 | 133 | Rochester University | US | 0,49 |
| 50 | 120 | 41 | 118 | Utrecht University | Netherlands | 0,47 |
| 51 | 109= | 69 | 271 | Case Western Reserve University | US | 0,43 |
| 52 | 19 | 82 | 81 | Melbourne University | Australia | 0,41 |
| 53 | 193 | 43 | 60 | Pittsburgh University | US | 0,32 |
| 54 | 45 | 71 | 217 | Heidelberg University | Germany | 0,30 |
| 55 | 66 | 57 | 240 | Copenhagen University | Denmark | 0,30 |
| 56 | 77= | 78 | 148 | Hebrew University of Jerusalem | Israel | 0,27 |
| 57 | 124 | 50 | 53 | University of Southern California | US | 0,27 |
| 58 | 62= | 76 | 63 | Helsinki University | Finland | 0,18 |
| 59 | 105= | 62 | 252 | Osaka University | Japan | 0,18 |
| 60 | 61 | 75 | 27 | Purdue University | US | 0,16 |
| 61 | 79 | 67 | 395 | Lomonosov Moscow State University | Russia | 0,09 |
| 62 | 24= | 93 | 543 | Ecole Normale Supérieure, Paris | France | 0,04 |
| 63 | 143= | 65 | 130 | Sheffield University | UK | -0,01 |
| 64 | 65 | 85 | 138 | Vienna University | Austria | -0,01 |
| 65 | 136= | 73 | 289 | Tohoku University | Japan | -0,04 |
| 66 | 68 | 101-152 | 75 | Massachusetts University | US | -0,06 |
| 67 | 138= | 72 | 245 | Leiden University | Netherlands | -0,07 |
| 68 | 54 | 80 | 67 | Boston University | US | -0,09 |
| 69 | 180= | 60 | 91 | Uppsala University | Sweden | -0,10 |
| 70 | 150= | 78 | 92 | Rice University | US | -0,15 |
| 71 | 121= | 77 | 26 | Michigan State University | US | -0,15 |
| 72 | 85 | 57 | 186 | Zurich University | Switzerland | -0,19 |
| 73 | 138= | 69 | 72 | Oslo University | Norway | -0,22 |
| 74 | 105= | 101-152 | 17 | Virginia University | US | -0,26 |
| 75 | 99 | 93 | 247 | Tokyo Institute of Technology | Japan | -0,28 |
| 76 | 71 | 86 | 87 | Brown University | US | -0,28 |
| 77 | 97 | 83 | 285 | Nottingham University | UK | -0,28 |
| 78 | 38= | 101-152 | 93 | Sydney University | Australia | -0,34 |
| 79 | 125= | 89 | 20 | Texas A&M University | US | -0,34 |
| 80 | 58= | 101-152 | 124 | Amsterdam University | Netherlands | -0,37 |
| 81 | 164 | 100 | 156 | Tufts University | US | -0,38 |
| 82 | 184= | 90 | 174 | McMaster University | Canada | -0,39 |
| 83 | 88= | 46 | 1675 | Pierre and Marie Curie University | France | -0,41 |
| 84 | 117 | 101-152 | 117 | Dartmouth College | US | -0,45 |
| 85 | 22 | 101-152 | 121 | National University of Singapore | Singapore | -0,47 |
| 86 | 73= | 80 | 377 | King's College London | UK | -0,47 |
| 87 | 143= | 98 | 129 | Birmingham University | UK | -0,49 |
| 88 | 147= | 101-152 | 40 | Georgia Institute of Technology | US | -0,51 |
| 89 | 188= | 101-152 | 280 | Tel Aviv University | Israel | -0,63 |
| 90 | 103= | 101-152 | 73 | Leeds University | UK | -0,65 |
| 91 | 142 | 101-152 | 204 | Frankfurt University | Germany | -0,68 |
| 92 | 180= | 99 | 197 | Lund University | Sweden | -0,71 |
| 93 | 15 | 203-300 | 213 | Beijing University | China | -0,72 |
| 94 | 119 | 101-152 | 303 | Liverpool University | UK | -0,73 |
| 95 | 141 | 101-152 | 127 | Emory University | US | -0,75 |
| 96 | 125= | 97 | 306 | La Sapienza University, Rome | Italy | -0,75 |
| 97 | 101= | 101-152 | 82 | Glasgow University | UK | -0,76 |
| 98 | 127= | 87 | 378 | Basel University | Switzerland | -0,76 |
| 99 | 194= | 101-152 | 183 | Technion - Israel Inst of Technology | Israel | -0,77 |
| 100 | 149 | 101-152 | 58 | University of Alberta | Canada | -0,80 |
| 101 | 129 | 101-152 | 192 | Nagoya University | Japan | -0,81 |
| 102 | 47 | 101-152 | 177 | Queensland University | Australia | -0,84 |
| 103 | 138= | 101-152 | 312 | Aarhus University | Denmark | -0,90 |
| 104 | 34 | 153-202 | 152 | Ecole Polytech Fédérale de Lausanne | Switzerland | -0,94 |
| 105 | 11= | 203-300 | 336 | London School of Economics | UK | -0,95 |
| 106 | 175= | 101-152 | 144 | State Univ of New York, Stony Brook | US | -0,97 |
| 107 | 93= | 101-152 | 316 | Seoul National University | South Korea | -1,01 |
| 108 | 40 | 153-202 | 125 | University of New South Wales | Australia | -1,03 |
| 109 | 95= | 153-202 | 149 | National Autonomous Univ of Mexico | Mexico | -1,07 |
| 110 | 88= | 101-152 | 98 | Geneva University | Switzerland | -1,09 |
| 111 | 57 | 153-202 | 625 | Erasmus University Rotterdam | Netherlands | -1,09 |
| 112 | 196= | 101-152 | 109 | São Paulo University | Brazil | -1,11 |
| 113 | 33 | 203-300 | 103 | Monash University | Australia | -1,14 |
| 114 | 132 | 153-202 | 101 | Université de Montréal | Canada | -1,19 |
| 115 | 95= | 101-152 | 161 | Catholic University of Leuven (Flemish) | Belgium | -1,24 |
| 116 | 179 | 153-202 | 111 | Notre Dame University | US | -1,26 |
| 117 | 157= | 101-152 | 362 | Hokkaido University | Japan | -1,28 |
| 118 | 191 | 203-300 | 116 | University of Western Ontario | Canada | -1,29 |
| 119 | 52 | 203-300 | 248 | Auckland University | New Zealand | -1,38 |
| 120 | 190 | 153-202 | 198 | Gothenburg University | Sweden | -1,39 |
| 121 | 83 | 203-300 | 267 | Durham University | UK | -1,40 |
| 122 | 80= | 153-202 | 230 | University of Western Australia | Australia | -1,41 |
| 123 | 10 | 203-300 | 801 | Ecole Polytechnique | France | -1,43 |
| 124 | 62= | 153-202 | 389 | Tsing Hua University | China | -1,44 |
| 125 | 114= | 153-202 | 170 | National Taiwan University | Taiwan | -1,44 |
| 126 | 41 | 203-300 | 295 | Hong Kong University | Hong Kong | -1,45 |
| 127 | 100 | 101-152 | 343 | Sussex University | UK | -1,51 |
| 128 | 154= | 153-202 | 268 | Technical University of Denmark | Denmark | -1,52 |
| 129 | 43 | 203-300 | 481 | Hong Kong University Sci & Technol | Hong Kong | -1,53 |
| 130 | 186= | 153-202 | 294 | Free University of Amsterdam | Netherlands | -1,55 |
| 131 | 168 | 203-300 | 139 | Newcastle upon Tyne University | UK | -1,58 |
| 132 | 159= | 203-300 | 172 | Georgetown University | US | -1,61 |
| 133 | 51 | 203-300 | 211 | Chinese University of Hong Kong | Hong Kong | -1,63 |
| 134 | 108 | 153-202 | 768 | Wageningen University | Netherlands | -1,64 |
| 135 | 183 | 153-202 | 375 | Madrid Autonomous University | Spain | -1,64 |
| 136 | 53 | 203-300 | 185 | Delft University of Technology | Netherlands | -1,67 |
| 137 | 154= | 203-300 | 100 | Technical University Berlin | Germany | -1,67 |
| 138 | 111 | 203-300 | 260 | Trinity College, Dublin | Ireland | -1,73 |
| 139 | 199= | 203-300 | 157 | George Washington University | US | -1,74 |
| 140 | 159= | 203-300 | 78 | Bologna University | Italy | -1,74 |
| 141 | 136= | 203-300 | 403 | St Andrews University | UK | -1,74 |
| 142 | 80= | 203-300 | 329 | Adelaide University | Australia | -1,77 |
| 143 | 77= | 203-300 | 195 | Warwick University | UK | -1,81 |
| 144 | 159= | 203-300 | 61 | University of Waterloo | Canada | -1,82 |
| 145 | 67 | 203-300 | 318 | Macquarie University | Australia | -1,83 |
| 146 | 199= | 203-300 | 171 | Wake Forest University | US | -1,83 |
| 147 | 172= | 203-300 | 135 | Aachen RWTH | Germany | -1,85 |
| 148 | 199= | 203-300 | 344 | University of Florence | Italy | -1,86 |
| 149 | 109= | 203-300 | 162 | York University | UK | -1,88 |
| 150 | 165 | 203-300 | 251 | Innsbruck University | Austria | -1,94 |
| 151 | 131 | 92 | 2797 | University Louis Pasteur Strasbourg | France | -1,96 |
| 152 | 130 | 301-400 | 278 | Bath University | UK | -1,99 |
| 153 | 166= | 203-300 | 146 | Chalmers University of Technology | Sweden | -2,03 |
| 154 | 196= | 203-300 | 120 | Royal Institute of Technology | Sweden | -2,05 |
| 155 | 147= | 203-300 | 445 | Hiroshima University | Japan | -2,05 |
| 156 | 172= | 203-300 | 513 | Kobe University | Japan | -2,12 |
| 157 | 143= | 301-400 | 455 | Korea Advanced Inst of Sci and Tech | South Korea | -2,15 |
| 158 | 177 | 203-300 | 720 | Nijmegen University | Netherlands | -2,17 |
| 159 | 86 | 301-400 | 94 | Vienna Technical University | Austria | -2,18 |
| 160 | 72 | 301-400 | 732 | Fudan University | China | -2,23 |
| 161 | 76 | 401-500 | 313 | Brussels Free University (French) | Belgium | -2,29 |
| 162 | 70 | 301-400 | 175 | Eindhoven University of Technology | Netherlands | -2,34 |
| 163 | 48 | 301-400 | 479 | Nanyang Technological University | Singapore | -2,37 |
| 164 | 169= | 301-400 | 687 | Shanghai Jiao Tong University | China | -2,43 |
| 165 | 186= | 301-400 | 490 | Otago University | New Zealand | -2,48 |
| 166 | 93= | 301-400 | 1009 | China University of Sci & Technology | China | -2,50 |
| 167 | 127= | 301-400 | 390 | University of Newcastle | Australia | -2,52 |
| 168 | 133= | 301-400 | 538 | Lausanne University | Switzerland | -2,54 |
| 169 | 150= | 301-400 | 852 | Nanjing University | China | -2,61 |
| 170 | 98 | 401-500 | 397 | La Trobe University | Australia | -2,71 |
| 171 | 166= | 401-500 | 359 | Tasmania University | Australia | -2,73 |
| 172 | 157= | 401-500 | 606 | Maastricht University | Netherlands | -2,74 |
| 173 | 178 | 301-400 | 434 | City University of Hong Kong | Hong Kong | -2,76 |
| 174 | 92 | 301-400 | 1032 | Ecole Normale Supérieure, Lyon | France | -2,80 |
| 175 | 154= | 401-500 | 510 | University of South Australia | Australia | -2,89 |
| 176 | 194= | 401-500 | 781 | Helsinki University of Technology | Finland | -2,97 |
| 177 | 184= | 401-500 | 930 | Korea University | South Korea | -3,02 |
| 178 | 188= | 401-500 | 575 | Massey University | New Zealand | -3,15 |

## Appendix B

### *Internationalism*

The extremes of Principal Component 2: The international component among faculty and student. The first ranked are institutions with large scores of foreign fraction among faculty or students.

| Rank PC-2 | Name | Country |
|---|---|---|
| 1 | London School of Economics | UK |
| 2 | Ecole Polytech Fédérale de Lausanne | Switzerland |
| 3 | Nanyang Technological University | Singapore |
| 4 | Geneva University | Switzerland |
| 5 | Cambridge University | UK |
| 6 | ETH Zurich | Switzerland |
| 7 | National University of Singapore | Singapore |
| 8 | Ecole Polytechnique | France |
| 9 | Oxford University | UK |
| 10 | Imperial College London | UK |
| ..... | ........................................................ | ................ |
| 169 | National Taiwan University | Taiwan |
| 170 | North Carolina University | US |
| 171 | University of California, Santa Barbara | US |
| 172 | Washington University | US |
| 173 | São Paulo University | Brazil |
| 174 | University of Minnesota | US |
| 175 | University of California, Los Angeles | US |
| 176 | University of California, San Diego | US |
| 177 | Pennsylvania State University | US |
| 178 | Tel Aviv University | Israel |

# Appendix C

## *Faculty/student ratio*

Extremes of projections of universities with respect to the Principal Component 3: The faculty to student ratio. The first ranked are institutions with high faculty/student ratios.

| Rank PC-3 | Name | Country |
|---|---|---|
| 1 | Ecole Polytechnique | France |
| 2 | University of California, San Francisco | US |
| 3 | University Louis Pasteur Strasbourg | France |
| 4 | Ecole Normale Supérieure, Paris | France |
| 5 | Ecole Normale Supérieure, Lyon | France |
| 6 | California Institute of Technology | US |
| 7 | Ecole Polytech Fédérale de Lausanne | Switzerland |
| 8 | Pierre and Marie Curie University | France |
| 9 | Eindhoven University of Technology | Netherlands |
| 10 | Duke University | US |
| ..... | ......................................................... | ................ |
| 169 | Nanyang Technological University | Singapore |
| 170 | Queensland University | Australia |
| 171 | University of British Columbia | Canada |
| 172 | London School of Economics | UK |
| 173 | University of New South Wales | Australia |
| 174 | University of Toronto | Canada |
| 175 | Sydney University | Australia |
| 176 | Melbourne University | Australia |
| 177 | Monash University | Australia |
| 178 | National University of Singapore | Singapore |